\documentclass[twocolumn,superscriptaddress,pra]{revtex4-1}
\usepackage{mathrsfs,braket}
\usepackage{amssymb, amsbsy, amsmath, latexsym, dsfont, array, layout,
graphicx,mathrsfs,color,ulem,bm}
\usepackage{float}
\usepackage{dcolumn}

\begin{document}

\date{\today}
\title{Impurity in a Fermi gas under non-Hermitian spin-orbit coupling}
\author{Jia-Zheng Sun}
\affiliation{CAS Key Laboratory of Quantum Information, University of Science and Technology of China, Hefei 230026, China}

\begin{abstract}
We study the fate of an impurity in a two-component, non-interacting Fermi gas under a non-Hermitian spin-orbit coupling (SOC) which is generated by dissipative Raman lasers. While SOC mixes the two spin species in the Fermi gas thus modifies the single-particle dispersions, we consider the case where the impurity only interacts with one of the spin species. As a result, spectral properties of the impurity constitute an ideal probe to the dissipative Fermi gas in the background. In particular, we show that dissipation destabilizes polarons in favor of molecular formation, consistent with previous few-body studies. The dissipative nature of the Fermi gas further leads to broadened peaks in the inverse radio-frequency spectra for both the attractive and repulsive polaron branches, which could serve as signals for experimental observation. Our results provides an exemplary scenario where the interplay of non-Hermiticity and interaction can be probed.
\end{abstract}

\maketitle

\section{Introduction}

Non-Hermitian systems have attracted much research interest lately due to their unique properties. In general,
the dynamics of a system can effectively be driven by a non-Hermitian Hamiltonian when it is coupled to an environment~\cite{QJ}. While some non-Hermitian systems still acquire purely real eigen spectra thanks to the parity-time symmetry~\cite{review1,review2}, some may possess properties such as skin effects and non-Bloch bulk-boundary correspondence~\cite{Lee,Budich,WZ1,WZ2,murakami,ThomalePRB,metaskin,photonskin,scienceskin}, with deep topological origins that are non-existent in their Hermitian counterparts~\cite{kawabataskin,fangchenskin}. A key issue in the study of non-Hermitian systems is the interplay of interaction and non-Hermiticity in a many-body setting~\cite{Ghatak,Zhou,Yamamoto,Ott4,Ott5,Durr,Takahashi1,Takahashi2,Wunner,Main,Konotop,Ueda1,Pan,Yu,Ueda2,zhaichen,Cui2020}. It has been shown that such an interplay gives rise to significantly modified dynamics in bosons~\cite{Durr,Takahashi1,Takahashi2}, as well as unconventional pairing superfluid in fermions~\cite{Zhou,Yamamoto}. In a very recent study, a non-Hermitian version of the spin-orbit coupling (SOC) has been proposed~\cite{Cui2020}, under which the SOC-dressed and dissipative single-particle dispersion results in a dissipation-facilitated molecular state. The impact of such a non-Hermitian synthetic gauge field in a genuine many-body system is still largely unexplored.

In this work, we bridge this knowledge gap by investigating an intermediate scenario interpolating the few- and many-body physics under a non-Hermitian SOC. Specifically, we study an impurity immersed in a two-component Fermi gas under the dissipative SOC. The two spin components of the Fermi gas are non-interacting by themselves, but the impurity has a tunable, spin-selective interaction with one of the spin species. Such a configuration is experimentally achievable by tuning close to a Feshbach resonance between the impurity atom and the relevant spin state of the Fermi gas. We focus on the parameter regime where the system is governed by a non-Hermitian Hamiltonian, which is satisfied when the evolution time is short compared to the inverse of the effective dissipation rate, i.e., when quantum jump terms are not dominant. Alternatively, by invoking a post-selection framework, fermions that still remain in the system are necessarily driven by a non-Hermitian Hamiltonian. Our model is therefore applicable for the duration when there are still appreciable fermions in the Fermi sea.

Adopting a Chevy-type anstaz, we then study the polaron state and molecules induced by the impurity-fermion interaction. Under the non-Hermitian SOC, both branches (attractive and repulsive) of polarons as well as the molecular states acquire complex energy spectra,  with the real and imaginary components corresponding to eigenenergies and widths, respectively. We show that by increasing the dissipative strength of the non-Hermitian SOC, the polaron-molecule transition point is shifted, from the BEC regime toward resonance. This observation is consistent with a previous study that suggests dissipation facilitates the formation of two-body bound states. We then apply the T-matrix formalism to characterize the inverse radio-frequency (r.f.) spectrum, where the shift and broadening of peaks under dissipation are clearly visible. In light of recent experimental progress in dissipative cold atomic gases~\cite{Ott1,Ott2,Ott3,Gerbier,Luo,Gadway}, our results would be helpful for future experimental studies.

Our work is organized as follows. In Sec.~II, we present our model and the single-particle dispersion. In Sec.~III, we show the complex energy spectra of attractive and repulsive polarons, as well as those of molecular states. We then calculate the inverse r.f. spectroscopy in Sec. IV. A summary is given in Sec. V.

\section{Model and single-particle dispersion}
\label{sec:model}

Following the derivation in Ref.~\cite{Cui2020}, a non-interacting, two-component Fermi gas under a non-Hermitian SOC is governed by the non-Hermitian Hamiltonian
\begin{align}
H_\text{f}=\sum_{\bf k} \Psi^{\dag}_{\bf k}h_{\text{f}}\Psi_{\bf k},\label{eq:H0}
\end{align}
where
\begin{align}
h_\text{f}=\begin{bmatrix}
\frac{\hbar^2}{2m}(\mathbf{k}+k_{0}\mathbf{e}_{x})^{2}-i\Gamma_{x} & \Omega-i\Gamma_{x} \\
\Omega-i\Gamma_{x} & \frac{\hbar^{2}}{2m}(\mathbf{k}-k_{0}\mathbf{e}_{x})^{2}-i\Gamma_{x}
\end{bmatrix},
\end{align}
and $\Psi_\mathbf{k}=[a_{\mathbf{k}\uparrow},a_{\mathbf{k}\downarrow}]^T$, with $a_{{\bf k}\sigma}$ the annihilation operator of fermions with momentum ${\bf k}$ and spin $\sigma$ ($\sigma=\uparrow,\downarrow$). Here, $\Omega-i\Gamma_x$ ($\Gamma_x>0$) is the SOC amplitude, $2\hbar k_0$ is the momentum transfer in the Raman process generating the SOC, $m$ is the atomic mass.

\begin{figure}[tbp]
\begin{center}
\includegraphics[width=0.5\textwidth]{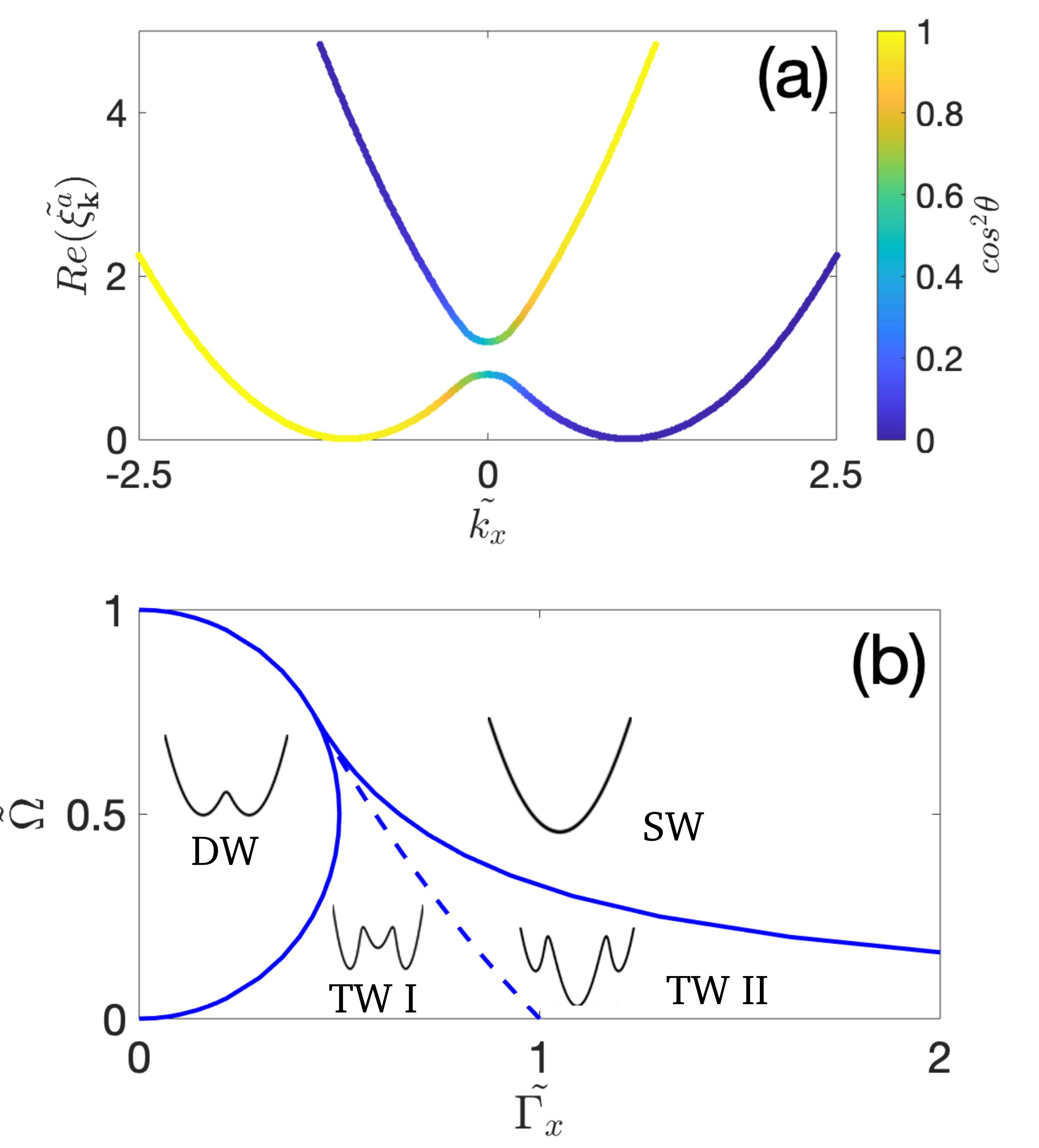}
\caption{(a) Real component of the single-particle dispersion under a non-Hermitian SOC for $\tilde{\Omega}=0.2$, $\tilde{\Gamma}_x=0.3$. Color code indicates the spin-up fraction. See main text for the definition of $\theta$. (b) Phase diagram for different geometry of the lower helicity band, with single-well (SW), double-well (DW) and two triple-well (TW I and TW II) regimes. 
We focus on the double-well (DW) regime, with a fixed $\tilde{k_0}=k_0/k_F=0.5$. Other dimensionless parameters are defined as $\tilde{\Omega}=\Omega/E_F$, $\tilde{\Gamma}_x=\Gamma_x/E_F$. The unit of energy $E_F$ and wave vector $k_F$ are defined in the main text.}
\label{fig:fig1}
\end{center}
\end{figure}

The single-particle dispersion of fermions $\xi_{\mathbf{k},\pm}$ can be derived by diagonalizing $h_f$, with
\begin{align}
\xi_{\mathbf{k},\pm}=
\frac{\hbar^{2}(k^{2}+k_{0}^{2})}{2m}-i\Gamma_{x}\pm
\sqrt{(\frac{\hbar^{2}k_{0}k_{x}}{m})^{2}+(\Omega-i\Gamma_{x})^{2}},
\end{align}
where $\pm$ labels the two helicity branches. The corresponding creation operators $a_{\mathbf{k},\pm}^{R\dag}$ for the right eigenstates of the helicity branches are given by
\begin{align}
a_{\mathbf{k},+}^{R\dag}&=
c_{1+}a_{\mathbf{k},\uparrow}^{\dag}+c_{2+}a_{\mathbf{k},\downarrow}^{\dag},\\
a_{\mathbf{k},-}^{R\dag}&=
c_{1-}a_{\mathbf{k},\uparrow}^{\dag}+c_{2-}a_{\mathbf{k},\downarrow}^{\dag},
\end{align}
where
\begin{align}
c_{1\pm}&=
\sqrt{\frac
{(\Omega^{2}+\Gamma_{x}^{2})\epsilon^\prime_{k}\cos\frac{\theta}{2}}
{\sqrt{\rho}[2\epsilon^\prime_{k}\sqrt{\rho}\cos\frac{\theta}{2}\pm(\Omega^{2}+\Gamma_{x}^{2}-\epsilon^{\prime 2}_{k}-\rho)]}},\\
c_{2\pm}&=
\frac{-\epsilon^\prime_{k}\pm\sqrt{\rho}e^{i\theta/2}}{\Omega-i\Gamma_{x}}
\times c_{1\pm}.
\end{align}
Here $\epsilon^\prime_k=\hbar^2k_0k_x/m$, $\rho=
\sqrt{(\epsilon^{\prime 2}_{k}+\Omega^{2}-\Gamma_{x}^{2})^{2}+4\Omega^{2}\Gamma_{x}^{2}}$, and $\tan \theta=-2\Omega \Gamma_{x}/(\epsilon^{\prime 2}_{k}+\Omega^{2}-\Gamma_{x}^{2})$.
In Fig.~\ref{fig:fig1}, we show the numerically evaluated single-particle dispersion, as well as the single-particle phase diagram. 
For numerical calculations, we take $E_F=\hbar^2k_F^2/2m$ as the unit of energy, where $k_F=(3\pi^2 n)^{1/3}$ and $n$ is the density of the Fermi sea. Throughout the work, we will focus on the parameter regime where a double-well (DW) structure exists in the lower helicity branch.

For the many-body setting, we assume that the Fermi energy is larger than the peak of the central barrier of the lower helicity branch, but smaller than the lowest energy of the higher helicity branch, so that only a single Fermi surface exists in the system. The actual Fermi energy $E_h$ is then a function of total density, as well as the SOC parameters~\cite{Cuipolaron}.
An impurity interacts with the spin-up component of the Fermi sea, with the Hamiltonian
\begin{align}
H_{\text{imp}}=\sum_{\mathbf{k}}
\epsilon_k b_\mathbf{k}^{\dag}b_\mathbf{k}+U/V\sum_{\mathbf{k,k',q}}
a_{\frac{\mathbf{q}}{2}+\mathbf{k},\uparrow}^{\dag}b_{\frac{\mathbf{q}}{2}-\mathbf{k}}^{\dag}
b_{\frac{\mathbf{q}}{2}-\mathbf{k'}}a_{\frac{\mathbf{q}}{2}+\mathbf{k'},\uparrow},
\end{align}
where $b_{\mathbf{k}}$ ($b^\dag_{\mathbf{k}}$) the annihilation  (creation) operator of the impurity atom, $V$ is the quantization volume, and $\epsilon_k=\hbar^2k^2/2m$ is the kinetic energy of the impurity atom, with the assumption that the impurity has the same mass as that of a fermion. The bare interaction strength $U$ is related to the $s$-wave scattering length $a_s$ between the impurity and the spin-up state through the standard renormalization relation in three dimensions $1/U=m/(4\pi \hbar^2 a_s)-1/V\sum_{\mathbf{k}}1/(2\epsilon_k)$.

\section{Polarons and molecules}

Following the Chevy's ansatz~\cite{impurityreview1,impurityreview2}, we write the polaron and molecular states as
\begin{align}
\vert P \rangle &=
(\phi_{0}b_{0}^{\dag}+\sum_{\lambda_{1},\mathbf{k}} \sum_{\lambda_{2},\mathbf{q}}
\phi_{\mathbf{k,q}}^{\lambda_{1},\lambda_{2}}b_{\mathbf{q-k}}^{\dag}
a_{\mathbf{k},\lambda_{1}}^{R\dag} a_{\mathbf{q},\lambda_{2}}^{L})
| \text{FS} \rangle_{N},\\
| M \rangle &=
\sum_{\lambda,\mathbf{k}}
\phi_{\mathbf{k}}^{\lambda} b_{\mathbf{-k}}^{\dag} a_{\mathbf{k},\lambda}^{R\dag}
| \text{FS} \rangle_{N-1},
\end{align}
where $\phi_0$, $\phi_{\mathbf{k,q}}^{\lambda_{1},\lambda_{2}}$, and $\phi_{\mathbf{k}}^{\lambda}$ are the corresponding wave functions, $|FS\rangle_N$ represents a Fermi sea with $N$ fermions.

From the Schr\"odinger's equations $(H_{\text{f}}+H_{\text{imp}})|P(M)\rangle=E_{P(M)}|P(M)\rangle$, we have the closed equations
\begin{align}
E_{P}&=\sum_{\lambda_{2}=\pm , \mathbf{q}} \frac
{\beta_{q}^{\lambda_{2} }}
{U^{-1}-
\sum_{\lambda_{1}=\pm , \mathbf{k}} \frac
{\beta_{k}^{\lambda_{1} }}
{E_{P}-\epsilon_{\mathbf{q-k}}-\xi_{\lambda_{1} , \mathbf{k}}+\xi_{\lambda_{2} , \mathbf{q}}}}\label{eq:pol}\\
\frac{1}{U}&=
\sum_{\lambda=\pm , \mathbf{k}} \frac
{c_{1\lambda}(\mathbf{k})\beta_{k}^{\lambda}}
{E_{M}-\xi_{\lambda , \mathbf{k}}^{a}-\epsilon_k}
(-1)^{\lambda},
\end{align}
where $\beta_{k}^{+}=(c_{1+}c_{2-})/(c_{1+}c_{2-}-c_{1-}c_{2+})$, $\beta_{k}^{-}=-(c_{1-}c_{2+})/(c_{1+}c_{2-}-c_{1-}c_{2+})$. While both $E_P$ and $E_M$ are complex, the ground state of the system, which is essentially a quasi-steady state in the short time scale, can be determined by comparing $\text{Re}(E_P)$ and $\text{Re}(E_M)-E_h$.

\begin{figure}[tbp]
\begin{center}
\includegraphics[width=0.48\textwidth]{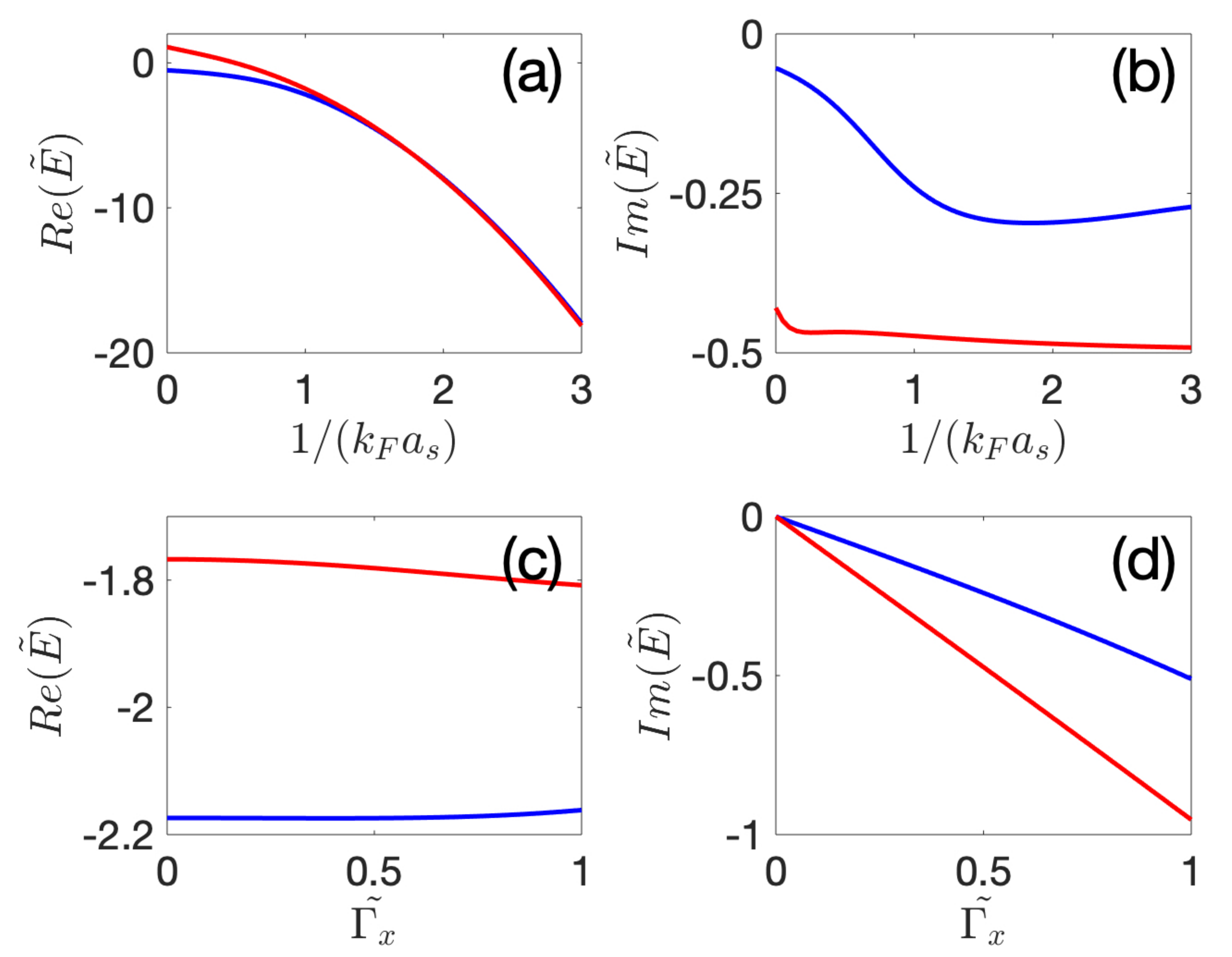}
\caption{(a) Real components of the attractive polaron $\text{Re}(E_P)$ (blue) and molecule $\text{Re}(E_M)-E_h$ (red) energies with varying interaction strength. (b) Imaginary components of the attractive polaron $\text{Im}(E_P)$ (blue) and molecule $\text{Im}(E_M)$ (red) energies.
(c) Real components of the attractive polaron $\text{Re}(E_P)$ (blue) and molecule $\text{Re}(E_M)-E_h$ (red) energies with increasing $\Gamma_x$. (d) Imaginary components of the attractive polaron $\text{Im}(E_P)$ (blue) and molecule $\text{Im}(E_M)$ (red) energies.
We set $\tilde{\Omega}=1$ and $\tilde{\Gamma}_x=0.5$ in (a)(b); $\tilde{\Omega}=1$ and $1/(k_Fa_s)=1$ in (c)(d). We fix $\tilde{k_0}=0.5$, and the dimensionless parameters are defined in the same way as those in Fig.~\ref{fig:fig1}.}
\label{fig:fig2}
\end{center}
\end{figure}

\begin{figure}[tbp]
\begin{center}
\includegraphics[width=0.48\textwidth]{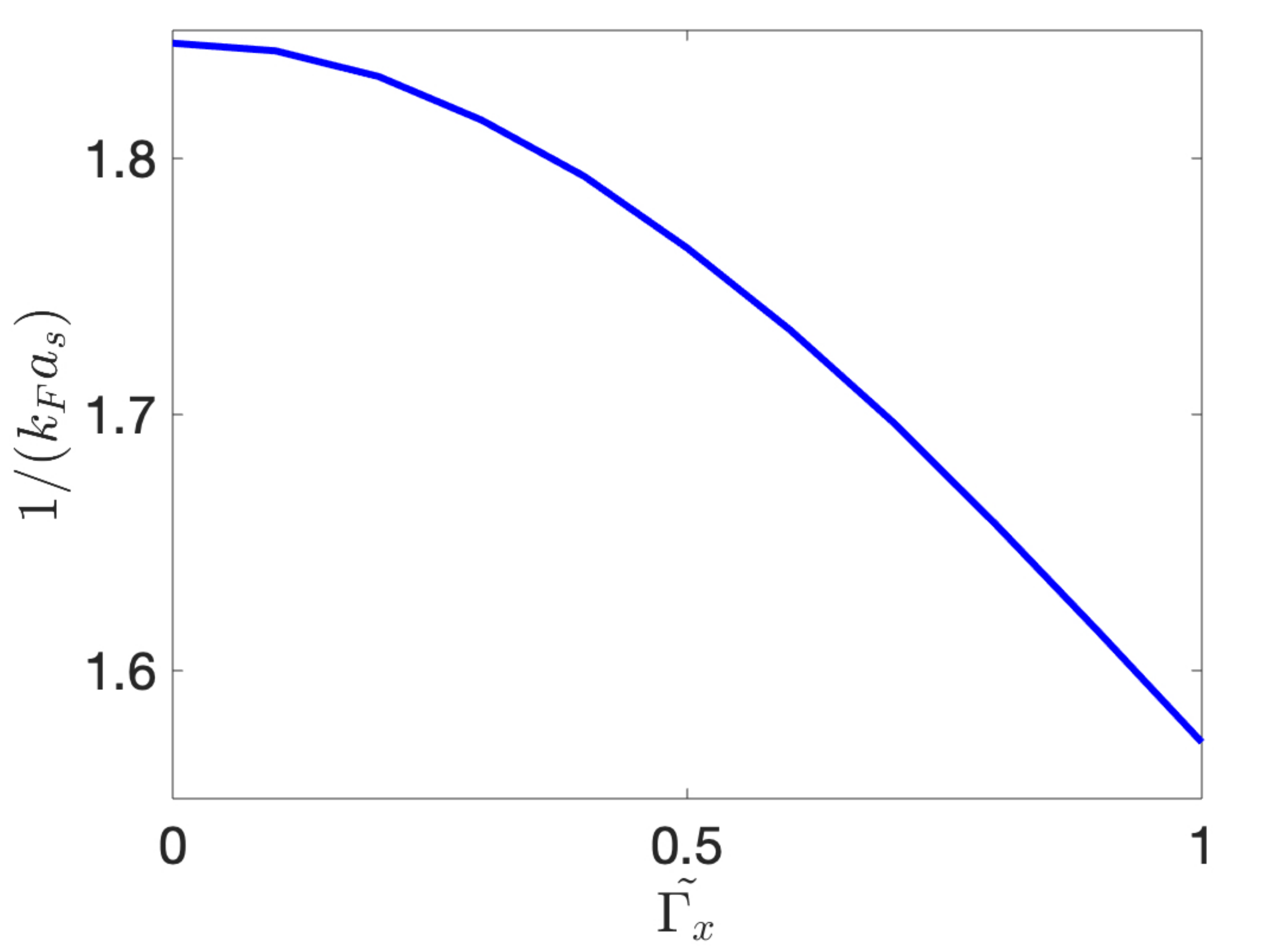}
\caption{Critical interaction strength for the polaron-molecule transition with increasing dissipation. We fix $\tilde{\Omega}=1$,  $\tilde{k_0}=0.5$, and the dimensionless parameters are defined in the same way as those in Fig.~\ref{fig:fig1}.}
\label{fig:fig3}
\end{center}
\end{figure}

In Fig.~\ref{fig:fig2}, we show the real and imaginary components of the polaron (blue) and molecular (red) energies, as functions of the interaction strength [(a)(b)] and dissipation [(c)(d)]. While both the polaron and molecule cease to be well-defined quasi-particles when the imaginary components (widths) of their energies become dominant over the real components, we find that they remain well-defined either deep in the BEC regime where interaction effects dominate over dissipation, or close to resonance but with small dissipation $\Gamma_x$. Further, a polaron-molecule transition can still be observed in Fig.~\ref{fig:fig2}(a). This is illustrated in detail in Fig.~\ref{fig:fig3}, where the polaron-molecule transition is found to shift toward the Feshbach resonance with increasing dissipation. This result is consistent with the previous finding that dissipation can facilitate molecular formation in vacuum.

\begin{figure}[tbp]
\begin{center}
\includegraphics[width=0.48\textwidth]{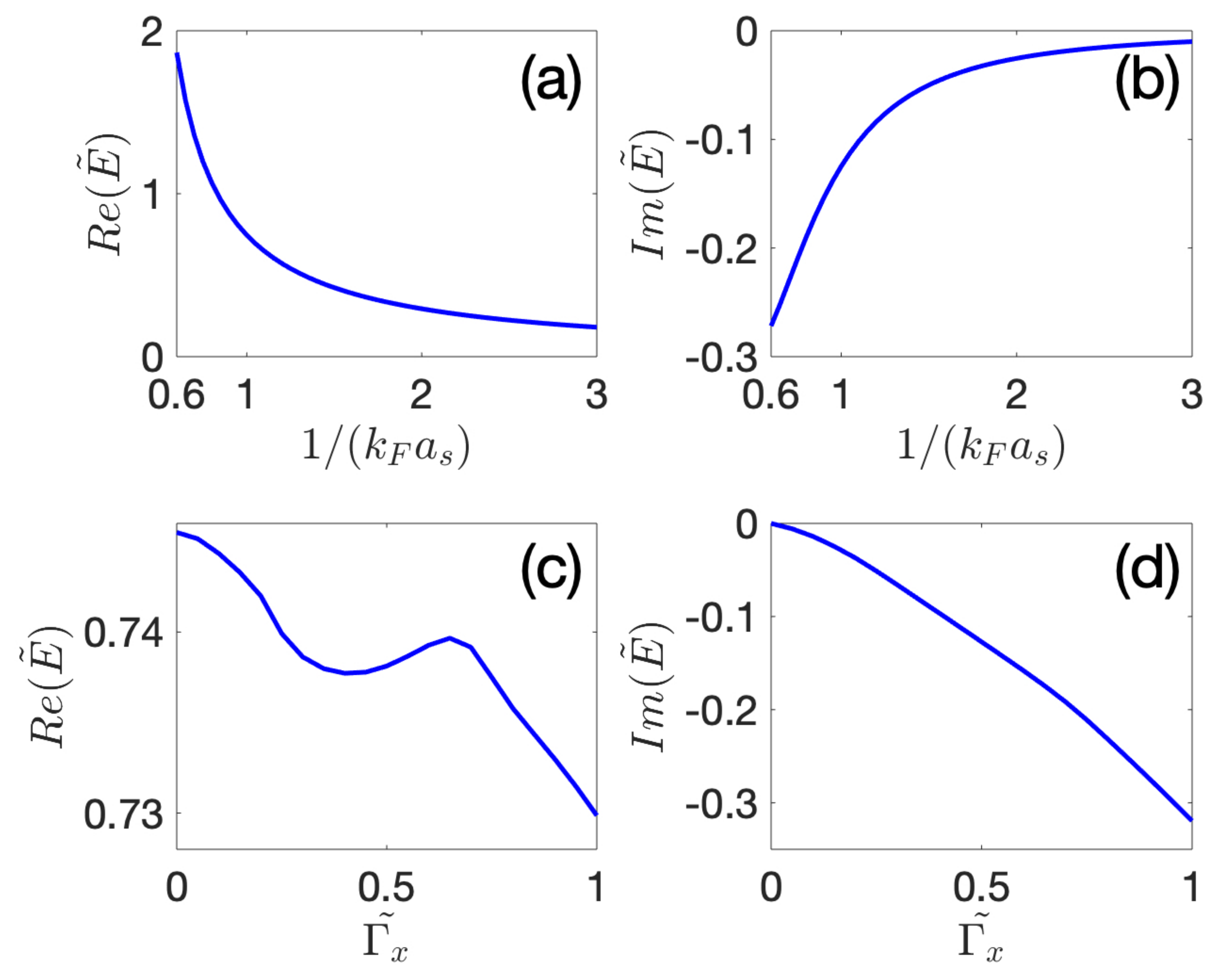}
\caption{Real (a)(c) and imaginary (b)(d) components of the repulsive-branch polaron energies.
 We take $\tilde{\Omega}=1$ and $\tilde{\Gamma}_x=0.5$ in (a)(b); and $\tilde{\Omega}=1$ and $1/(k_Fa_s)=1$ in (c)(d). We also fixe $\tilde{k_0}=0.5$.}
\label{fig:fig4}
\end{center}
\end{figure}

Equation (\ref{eq:pol}) also allows us to solve for polarons of the repulsive branch. This is shown in Fig.~\ref{fig:fig4}, where we demonstrate how the real and imaginary components of the repulsive polaron energy vary with interaction [(a)(b)] and dissipation [(c)(d)]. Under a fixed dissipation strength, the imaginary components of the repuslive polaron energy decrease rapidly toward the BEC regime. This originates from the fact that the attractive polaron branch becomes much lower in energy, and that dissipation effect from the non-Hermitian SOC becomes suppressed in the strong-interaction regime.
Therefore, despite a decreasing tendency of the polaron energy in the BEC regime, the repulsive branch remains well-defined. On the other hand, when dissipation becomes sufficiently large, the width of the repulsive branch should become appreciable, and the polaron description would fail.

\section{Detecting polarons: r.f. spectroscopy}

While the impurity serves as a probe for the non-Hermitian Fermi sea, it can leave key signatures in the r.f. spectroscopy. We focus on the inverse r.f. spectroscopy here, where we prepare the impurity atom in a bystander state that does not interact with the Fermi sea, and then couple the bystander state to the actual impurity state that interacts with the spin-up fermions. We note that such a scheme does not require a detailed protocol on preparing the quasi-steady polaron state: it can be resolved so long as it is a well-defined quasi-particle with the imaginary component of its energy much smaller than the real component.

Following the linear-response theory~\cite{Cui2020}, the population transfer or the signal of the r.f. spectroscopy is given by
\begin{align}
R(\omega)=2\text{Im}\frac{1}{\omega-\Sigma(0,\omega)},
\end{align}
where the self-energy is given by
\begin{align}
\Sigma(0,\omega)=
\sum_{\lambda_{2} , \mathbf{q}} \frac
{\beta_{\mathbf{q}}^{\lambda_{2} }}
{U^{-1}-
\sum_{\lambda_{1} , \mathbf{k}} \frac
{\beta_{\mathbf{k}}^{\lambda_{1} }}
{\omega-\epsilon_{\mathbf{q-k}}^{b}-\xi_{\lambda_{1} , \mathbf{k}}^{a}+\xi_{\lambda_{2} , \mathbf{q}}^{a}}}. \label{eq:self}
\end{align}

\begin{figure}[tbp]
\begin{center}
\includegraphics[width=0.48\textwidth]{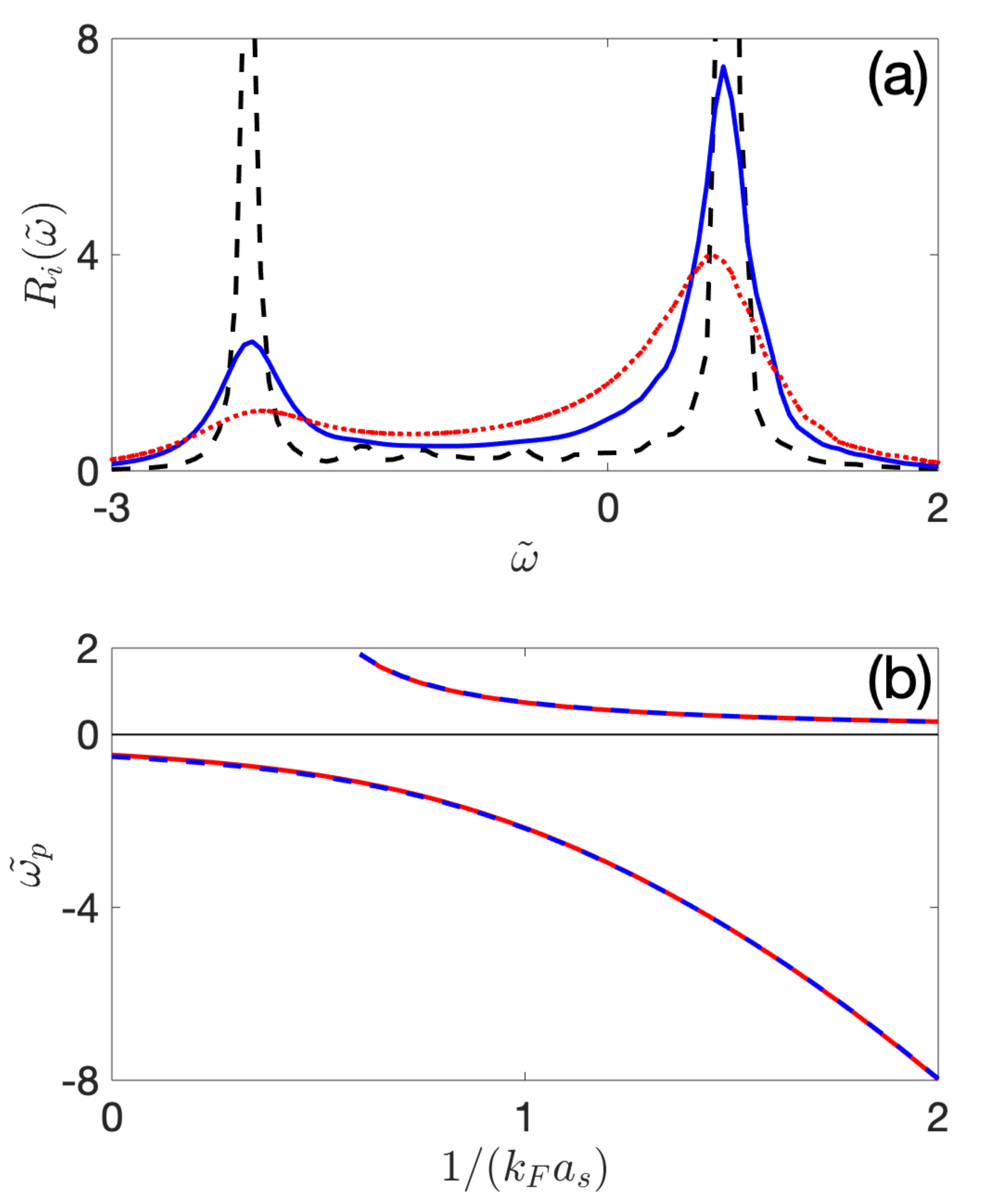}
\caption{(a) Inverse r.f. spectra $R(\omega)$ for $\tilde{\Gamma}_x=0.1$ (black dashed), $\tilde{\Gamma}_x=0.5$ (blue solid), and $\tilde{\Gamma}_x=1$ (red dash-dotted). We take $\tilde{\Omega}=1$, $1/(k_Fa_s)=1$ for the calculation. (b) Comparison of the real component of polaron energies from the variational calculations Eq.~(\ref{eq:pol}) (blue) and the peaks in the r.f. spectrum (red dashed). We take $\tilde{\Omega}=1$, $\tilde{\Gamma}_x=0.5$, and $\tilde{k_0}=0.5$ here.}
\label{fig:fig5}
\end{center}
\end{figure}

In Fig.~\ref{fig:fig5}(a), we show the calculated inverse r.f. spectra with different dissipation strength $\Gamma_x$. Both attractive and repulsive polaron branches are clearly visible as peaks in the spectrum, where the widths of peaks increase for larger $\Gamma_x$.
Note that compared with Eq.~(\ref{eq:pol}), $\omega$ on the right-hand side of Eq.~(\ref{eq:self}) is real. Therefore, under the non-Hermitian SOC, peaks in the inverse r.f. spectrum does not exactly correspond to $\text{Re}(E_P)$ calculated from Eq.~(\ref{eq:pol}). This is in sharp contrast to the Hermitian case. However, as we illustrate in Fig.~\ref{fig:fig5}(b), positions of the spectral peaks do no deviate much from polaron energies calculated using the variational approach. Hence, the dissipative polarons can be fully resolved by the inverse r.f. spectroscopy.

\section{Summary}

In this work, we study the quasi-particles induced by an impurity in a two-component Fermi gas under a non-Hermitian SOC. We show dissipation affects the polaron-molecule transition in the system, stabilizing the molecular state. We further demonstrate that both attractive and repulsive polarons can be resolved by the inverse r.f. spectroscopy, thus providing an ideal experimental probe to the system. Our configuration has highly tunable parameters: the dissipative SOC is directly tunable through laser parameters of the Raman process; the interaction is tunably through the Feshbach resonance; the Fermi surface can be adjusted by controlling the fermion densities as well as the SOC parameters. Further, our detection scheme using the inverse r.f. spectroscopy does not require a polaron state to exist in the dissipative system prior to detection, thus circumventing a key difficulty in non-Hermitian many-body systems where stringent requirements on the hierarchy of time scales may hinder experimental preparation and detection. Our work therefore provides an ideal scenario for the investigation of non-Hermitian many-body systems.

\section*{Acknowledgements}

We thank Lihong Zhou, Jing Zhou, Xiaoling Cui, and Wei Yi for helpful discussions. 

\end{document}